# Quantum patterns of genome size variation in angiosperms

Liaofu Luo[1,2]     Lirong Zhang[1]

[1] School of Physical Science and Technology, Inner Mongolia University, Hohhot, Inner Mongolia, 010021, P. R. China

[2] School of Life Science and Technology, Inner Mongolia University of Science and Technology, Baotou, Inner Mongolia, 014010, P. R. China

Email to LZ: pyzlr@imu.edu.cn ;    LL: lolfcm@imu.edu.cn

**Abstract**

The nuclear DNA amount in angiosperms is studied from the eigen-value equation of the genome evolution operator ***H***. The operator ***H*** is introduced by physical simulation and it is defined as a function of the genome size *N* and the derivative with respective to the size. The discontinuity of DNA size distribution and its synergetic occurrence in related angiosperms species are successfully deduced from the solution of the equation. The results agree well with the existing experimental data of Aloe, Clarkia, Nicotiana, Lathyrus, Allium and other genera. It may indicate that the evolutionary constrains on angiosperm genome are essentially of quantum origin.

**Key words:** Genome size; DNA amount; Evolution; Angiosperms; Quantum

**Introduction**

DNA amount is relatively constant and tends to be highly characteristic for a species. The nuclear DNA amounts of angiosperm plants were estimated by at least eight different techniques and mainly (over 96%) by flow cytometry and Feulgen microdensitometry. The C-values for more than 6000 angiosperm species were reported till 2011(1). It is found that these C-values are highly variable, differing over 1000-fold. Changes in DNA amount in evolution often appear non-random in amount and distribution (2-4). One big surprise is the discovery of the large differences in DNA amounts between related plant species, while preserving the strictly proportional change in all chromosomes (2, 3, 5). As early as in 1980's, Narayan discovered the discontinuous DNA distribution in four angiosperm genera and named it as "the quantum mode of C-value progression" (6, 7). The discontinuous distribution of nuclear DNA amounts in angiosperm genera is rather common. It was also observed in genus Rosa (8). Both mechanisms on genome size increase and on genome size reduction in plant evolution were proposed (2, 3, 9-12). It was suggested that in plants the genome sizes increase due to the proliferation of long terminal repeats (LTR) retrotransposons (9). While the mechanism of genome reduction may be due to many small deletions (10) or some unknown force removing LTR retrotransposons and excess DNA (11). However, all these works cannot explain the law on the quantum pattern of the discontinuous DNA distribution in angiosperms. In the manuscript we shall introduce a genome evolution operator ***H***, called Hamiltonian in physics, and use the eigen-functions of the operator to deduce the distribution of DNA amount for a species or genus. Therefore, the law obeyed by the statistical distribution of genome size is similar to the eigen-equation of observables in quantum mechanics and this equation can easily be generalized to the time-dependent form and used as a guard for studying the plant genome evolution.

## Results

**Genome evolution operator** $H(N, \frac{\partial}{\partial N})$   To describe the discontinuous patterns of genome size distribution we simulate the genome evolution by a physical system. The system is characterized by a genome evolution operator $H(N, \frac{\partial}{\partial N})$, where $N$ is the genome size. We assume the eigenvalue and the eigen-function of $H$ describe the organization of DNA amounts. Assume

$$H = -2(\frac{L}{c})^2 \frac{\partial^2}{\partial N^2} - V(N) = -2(\frac{L}{c})^2 \frac{\partial^2}{\partial N^2} - (2+\alpha)N + \beta N^2 \qquad (1)$$

The Hamiltonian includes the "kinetic" term $-2(\frac{L}{c})^2 \frac{\partial^2}{\partial N^2}$ and the "potential" term $V(N) = (2+\alpha)N - \beta N^2$. $L/c$ is an evolution-related constant for a family or a genus of species. $\alpha$ and $\beta$ are environment-related parameters. Without loss of generality one may assume $\alpha$ much smaller than 2. The eigenvalue $E$ and eigen-function $\psi(N)$ obey the Schrodinger equation

$$H\psi(N) = E\psi(N) \qquad (2)$$

that describes the steady states of the chromosome organization of the evolving species. The solution of Eq(2) is generally discontinuous. The discrete eigenvalues of Hamiltonian characterize the quantum states of the genome size and the corresponding eigen-function gives the distribution of genome size, namely $|\psi(N)|^2 dN$ representing the occurring probability of size $N$ in the state $\psi(N)$. The probability distribution may peak at one value or several values. As the probability peaks at two or more values it means the nuclear DNA amounts varying in that species.

The solutions of the eigenvalue equation (2) or its explicit form

$$\{\frac{-2L^2}{c^2}\frac{\partial^2}{\partial N^2} - (2+\alpha)N + \beta N^2\}\psi(N) = E\psi(N) \qquad (3)$$

are

$$E = E_n = (n+\frac{1}{2})\sqrt{8\beta}\frac{L}{c} - \frac{(2+\alpha)^2}{4\beta} \quad (n = 0, 1, 2, ...) \qquad (4)$$

and

$$\psi(N) = \psi_n(N) = C_n \exp(-\frac{\xi^2}{2})H_n(\xi)$$
$$H_0(\xi) = 1, \quad H_1(\xi) = 2\xi, \quad H_2(\xi) = 4\xi^2 - 2, \quad ...$$
$$H_{n+1}(\xi) = 2\xi H_n(\xi) - 2nH_{n-1}(\xi) \qquad (5)$$
$$\xi = \frac{1}{\sqrt{d}}(N - N_0) = \sqrt{\frac{c}{L}}(\frac{\beta}{2})^{1/4}(N - \frac{2+\alpha}{2\beta})$$
$$N_0 = \frac{2+\alpha}{2\beta}, \quad d = \frac{L}{c}\sqrt{\frac{2}{\beta}}$$

$C_n$ is normalization constant. The function $H_n(\xi)$ is called Hermite polynomial in mathematics. It is easily shown that the probability distribution of ground state (*n*=0) peaks at $\xi = 0$ or $N = N_0 = (2+\alpha)/2\beta$, the probability of the first excited state (*n*=1) having two peaks at $\xi = 1$ and -1, and the probability of the second excited state (*n*=2) having three peaks at $\xi = \sqrt{5/2}$, 0 and $-\sqrt{5/2}$, etc.

**Discontinuous distribution of DNA amounts of long and short chromosomes in Aloe** The Aloaceae is one of the most stable angiosperm families as far as gross chromosome morphology and basic number are concerned. The genome size of all species is bimodal, with several long chromosomes and several short ones in the diploid. The 4C DNA amounts were studied for 20 Aloe species and it was found the overall DNA amounts can be divided into long and short chromosome categories. The ratios of DNA amount between long and short chromosome are very close to 4.6:1 for 20 species (5). This provides a striking example of karyotypic orthoselection but its mechanism is unknown. We shall explain it as the quantum pattern of genome size. Suppose the genomes of the studied 20 Aloe species are all in the first excited states, *n*=1. From Eq (5) one has the probability density of genome size of the first excited state

$$|\psi_{n=1}(N)|^2 = const. \times \frac{4}{d}(N-N_0)^2 \exp[-\frac{(N-N_0)^2}{d}] \qquad (6)$$

It follows that the probability density peaks at

$$\begin{aligned} N = N_+ = N_0 + \sqrt{d} \quad &(for \quad long \quad chromosomes) \\ N = N_- = N_0 - \sqrt{d} \quad &(for \quad short \quad chromosomes) \end{aligned} \qquad (7)$$

Therefore one has

$$\frac{N_+}{N_{tot}} = \frac{1}{2} + \frac{\sqrt{d}}{2N_0}, \quad \frac{N_-}{N_{tot}} = \frac{1}{2} - \frac{\sqrt{d}}{2N_0} \quad (N_{tot} = N_+ + N_-) \qquad (7.1)$$

where $N_0 = N_{tot}/2$, $\sqrt{d} = (N_+ - N_-)/2$. The two parameters have been defined by Eq(5). Note that $N_0$ is also the DNA amount or genome size of the ground state $\psi_{n=0}$ and $d$ means the width of the size distribution in ground state. Eq (7.1) gives the DNA amount ratio between long and short chromosomes of *n*=1 state. From Eq (5) one has

$$\frac{\sqrt{d}}{N_0} = \frac{2^{5/4}\sqrt{L/c}\beta^{3/4}}{2+\alpha}$$

which is a constant for Aloe family. Therefore the theory predicts the DNA amounts of long and short DNA fall into two straight lines in the figure of $N_+$ and $N_-$ *versus* $N_{tot}$. The comparisons between theory and experiments are shown in Figure 1. We found the theory of nuclear DNA distribution is in good agreement with the experimental data. Moreover, the experimental data $N_+/N_- = 4.6$ gives

$d = 0.413N_0^2$ that implies the relation of parameters $L/c = 0.292\beta^{-3/2}$ (as assuming $\alpha$ much smaller than 2) for Aloeceae family.

[INSERT FIGURE 1 HERE]

**Discontinuous distribution of nuclear DNA amount in four genera Clarkia, Nicotiana, Lathyrus and Allium**   The discontinuous distribution of nuclear DNA amount also occurs in other angiosperms. The 2C DNA values of 125 species divided into four genera (Clarkia, Nicotiana, Lathyrus and Allium) were studied and they show the same discontinuous distribution of 4-8 groups (4 groups for Clarkia, 7 groups for Nicotiana and 8 groups for Lathyrus and Allium). It was reported that the group means occur at regular intervals averaging approximately 2pg in Clarkia and Nicotiana and approximately 4 pg in Lathyrus and Allium (7). The above empirical laws can be explained naturally in the present theory.

[INSERT TABLE 1 HERE]

Set the distribution of genome size in quantum state $n$ peaks at $\xi_i^{(n)}$ ($i=1,..,n+1$). The expression of wave function $\psi_n(N)$ in Eq (5) shows the DNA amount $N$ in quantum state $n$ always peaks at

$$N_i^{(n)} = N_0 + \sqrt{d}\,\xi_i^{(n)}$$
$$N_i^{(n)} - N_{i+1}^{(n)} = \sqrt{d}\,(\xi_i^{(n)} - \xi_{i+1}^{(n)}) \quad (8)$$

Since $\xi_i^{(n)}$ is discrete in given quantum state $n$, the genome size always shows nodal distribution.

From the eigen-function $\psi_n(N)$ given by Eq (5) the positions of peaks of genome size are determined by an algebraic equation of parameter $\xi$. The equations for $n=0$ to $n=7$ and their roots $\xi_i^{(n)}$ are listed in Table 1. Since the adjacent roots $\xi_i^{(n)}$ and $\xi_{i+1}^{(n)}$ depart each other by a number of 1 to 2 approximately one has $N_i^{(n)} - N_{i+1}^{(n)}$ approximately equal $\sqrt{d}$ to $2\sqrt{d}$ for each quantum state $n$. Because $d$ is a constant for a given genus the group means of DNA amount should occur at regular intervals of $\sqrt{d}$ to $2\sqrt{d}$. Generally, the theoretical value of nuclear DNA amount $N_i^{(n)}$ can be predicted by Eq (8) through two parameters $N_0$ and $\sqrt{d}$. Two parameters are estimated by

$$N_0 = \frac{1}{n+1}\sum_i N_i^{(n)} \quad (9)$$

$$(\sqrt{d})_1 = \left\langle \frac{N_i^{(n)} - N_{i+1}^{(n)}}{\xi_i^{(n)} - \xi_{i+1}^{(n)}} \right\rangle_{AV} \quad (10)$$

where $\langle \ \rangle_{AV}$ means the average over $i$. The parameters $N_0$ and $\sqrt{d}$ can also be estimated through minimization of the deviation

$$\sigma^2 = \frac{1}{n_{terms}} \sum_i [N_i^{(n)} - N_i^{(n)}(\exp)]^2$$

Here $N_i^{(n)}(\exp)$ represents the experiment value of $N_i^{(n)}$ (7) and $n_{terms}$ means the number of predicted species. For given $n$, the minimization leads to the estimation Eq (9) and

$$(\sqrt{d})_2 = \frac{\sum_i \{[N_i^{(n)}(\exp) - N_0]\xi_i^{(n)}\}}{\sum_i (\xi_i^{(n)})^2} \qquad (11)$$

Assume the nuclear DNA distribution of Clarkia is in the excited quantum state of $n=3$, Nicotiana in quantum state $n=6$, Lathyrus in quantum state $n=7$ and Allium in quantum state $n=7$. The comparisons between theoretical prediction and experimental data of nuclear DNA amounts for four genera are shown in Table 2.

[INSERT TABLE 2 AND TABLE 2a HERE]

From Table 2 we found the theoretical predictions $N_i^{(n)}(th1)$ and $N_i^{(n)}(th2)$ agree well with experimental data if the possible errors existing in group means of 2C DNA (see ref (7)) are taken into account. The studied genera, Clarkia, Nicotiana and Lathyrus, have complete DNA amount data. The standard deviation $\sigma$ for these three genera takes 0.191, 0.335 and 1.076 respectively from our quantum pattern model. For comparison we consider another measure by introducing empirical spacing parameter $\xi^{(n)} = \frac{N_1^{(n)} - N_{n+1}^{(n)}}{n}$ and use the equal spacing model to simulate the experimental data of these genera. The standard deviation takes 0.419, 0.415 and 1.42 respectively, explicitly higher than the deviations calculated from the quantum pattern model (Table 2a). The nuclear DNA distribution is a property fundamental to the complex organization of angiosperm genomes. Above studies proves that in these four genera the amounts of DNA form regular "nodes" along the entire sequence. The agreement between theory and experiments shows that the quantum mechanism governing the discontinuous pattern of DNA distribution is an acceptable assumption and this mechanism may be common to many plant genera.

The parameters $N_0$ and $(\sqrt{d})$ are related to more fundamental evolutionary parameters $L/c$ and $\beta$. The $\beta$ value reflects the environmental influence on the genus and $L/c$ reflects the inherent characteristics of the genus. From the values of $N_0$ and $(\sqrt{d})_2$ listed in Table 2, by using Eq (5) one obtains $L/c$ ranging from 0.7 to 2.8 pg$^{3/2}$ and $\beta$ from 0.03 to 0.19 pg$^{-1}$ for four genera.

The evolutionary parameters of these four plant genera take roughly similar values. It provides an explanation why their nuclear DNA amounts show the nearly same discontinuous distribution.

**Discontinuous nuclear DNA amount and quantum state in Rosa**    The above method can be used to study discontinuous distributions of nuclear DNA amounts in other species of angiosperms. The DNA amount can always be expressed by quantum states $S_i^{(n)}$ ($i$=1, 2,…$n$+1) where $n$=0 means ground state and $n \geq 1$ means excited state. As an example we study nuclear DNA amounts in roses. The 2C DNA amounts of rose species for four subgenera were estimated by using flow cytometry and Feulgen microdensitometry as summarized in ref 8 (8). Three subgenera have only one species and each of them should be classified as the quantum ground state $S_1^{(0)}$. The fourth subgenus Eurosa contains 120 species grouped into ten sections. Several discrete values of the 2C DNA amounts were taken in section Carolinae, Cinamomeae and Pimpinellifoliae. We define the nuclear quantum states of all species in subgenus Eurosa in Table 3 . The distribution parameters of DNA amounts $N_0$ , $\sqrt{d}$ and the parameter $\frac{L}{c}\beta^{3/2}$ (see **discussion** section) are also calculated for each quantum state and given in Table 3.

We have found quantum states $S_i^{(n)}$ of $n$=3, 6, 7 in Clarkia, Nicotiana, Lathyrus and Allium and $n$=1 in Aloe. Now the quantum states $S_i^{(n)}$ of $n$=0, 1 and 2 are found in genus Rosa. It is expected that all low-excited quantum states can be found in the genera of angiosperms. In addition to the discreteness of the eigenvalue (discontinuous pattern) the peculiarity of a quantum state is the statistical distribution of the genome size which possibly takes several different values following statistical law. The probability of genome size taking $N$ to $N$ +d$N$ is proportional to $|\psi(N)|^2 dN$. By use of Eq (5) we can deduce the approximately same probabilities at different peaks of $|\psi(N)|^2$.

[INSERT TABLE 3 HERE]

**Discussions**

**Two phases of genome evolution and the mechanical simulation of evolutionary theory**    The nuclear genomes in angiosperms exhibit extensively variation in size. The size variation has become an important part of the problem of genome macroevolution. Paleobiological studies indicated that species usually change more rapidly during, rather than between, speciation events. The smooth evolution always occurs between speciation events and the sudden evolution preferably occurs during speciation (13). We call the former as the evolution in classical phase because the phyletic gradualism obeys the classical evolutionary law. While the fossil record does not always show smooth evolutionary transitions. Punctuated equilibrium states mean that evolution is fast at times of splitting (speciation) and comes to a halt (stasis) between splits (14). In fact, the creation of new gene loci with previously nonexistent functions during speciation requires the big and rapid leap in genome evolution, and that is difficult to be understood by the classical evolution theory merely based on natural selection. Due to the abrupt and stochastic nature of speciation events we attribute the evolution during speciation to the non-classical or the quantum phase.

Alternative occurrence of classical and non-classical phase can be explained in a unifying evolutionary theory as follows. Suppose the frequency of aptamer ($k$-mer) serves the dynamical variables of the genome (15). Set the frequency $\{x_i\}$ ($i=1,...,m=4^k$) and $\sum x_i = N$ (genome length or size) are the function of evolutionary time $t$. They obey a set of second-order differential equations with respect to $t$. By use of mechanical simulation one may introduce kinetic energy, potential and Hamiltonian of the system. The advantage of the evolution expressed in the form of second-order differential equations is that it can easily be transformed into a quantum equation (see Supplementary data A, arXiv:1411,2205). When the evolutionary inertia parameter $c^2$ in equations is small enough during speciation, the classical mechanical equations are automatically transformed into the quantum Schrodinger equation that can describe the genome evolution in quantum phase. The approach was applied in the discussion of avian genome evolution. It was found that as $c$ decreases to the thousandth of the classical value the avian genome evolution enters into quantum phase (see Supplementary data B, bioRxiv 2015/034710).

More than 6000 angiosperm species have been studied and their DNA amounts are highly variable differing over 1000-fold. The species diversification is a process of consecutively alternative occurrence of classical and quantum phases of evolution. During speciation the genome evolution is in quantum phase denoted by $n$ with $n=0$ representing DNA amount singlet and $n \geq 1$ representing DNA amount multiplet. Then, after speciation the evolution enters to the classical phase where the component $S_i^{(n)}$ ($i=1, 2,...n+1$) in each given $i$ evolves as a new species moves on classical trajectory until next round of speciation.

**Genome size variations**   It was proposed that the genome sizes increase in angiosperms is due to the proliferation of long terminal repeat (LTR) retrotransposons, and the accumulation of several thousand new transposable elements in the long term of evolution (9). Simultaneously, it was suggested that the DNA loss from large-scale rearrangements is due to hundreds of thousands of small deletions in noncoding DNA and transposons (10, 11). But the transposable elements proliferation and the deletion-based mutation cannot explain the genome complexity observed in discontinuous DNA distribution in angiosperms. Recently the Whole Genome Duplication (WGD) as a major driving force in angiosperms and other species diversification is recognized (16-18). It was found that several rounds of WGD form cycles of polyploidization and the increased diversification rates often follow WGDs (19, 20). However, the polyploidy is often described as a somatic event of chromosome doubling and the mechanisms of chromosome doubling is unclear yet (21). In terms of polyploidization we are still unable to demonstrate how the relation of discontinuous DNA size distribution in related species is established. We noticed that the large-scale gene duplications (block duplications), both on the same chromosome and on different chromosomes, were studied experimentally and it was demonstrated that these events mainly result from replication accidents (22). One may assume the accidents have something to do with quantum law. In this paper we suggest a new way to solve the problem of discontinuous DNA distribution in angiosperms. By use of the eigenvalues of the genome evolution operator $H$ we have deduced the quantum pattern of genome size variation consistent with experimental data. This is the first success of the present method. Secondly, since $L/c$ and $\beta$ are two fundamental constant introduced in the operator $H$, the quantum pattern of nuclear DNA amount should depend on these two parameters. The discontinuous DNA amounts in related plant species should obey the law given by similar $L/c$ and $\beta$ and they occur synergetically. This is the very reason why the large differences in DNA amounts between related plant species preserve the strict proportionality such as observed in Aloe. To summarize, the genome size variations have two remarkable characteristics, the discontinuity of DNA size distribution and its synergetic occurrence in related angiosperms species. Both they can be understood and calculated following the line of the present theory. Moreover, the events of new species formation occur suddenly and stochastically. The puzzle can also be explained by quantum evolution.

**Dimension of evolutionary parameters** The dimension analysis is helpful to understand the meaning of the parameter introduced in the theory. The dimension of dynamical variable $N$ is zero since $N$ is measured by base number. However, the genome size $N$ can also be given as the mass of DNA (in picogram, pg, 1pg=0.98 $\times 10^9$ base pairs) (2, 3). So under the alternative definition, $N$ has dimension of mass [M]. That is, there are two dimensional systems of genomic quantity and both systems can be used in theoretical analysis. Genome evolution operator $H$ and its eigenvalue $E$ describe the discontinuous pattern of genome size, having the dimension as genome size $N$, namely dimension zero in the first dimensional system or [M] in the second dimensional system. By the physical simulation $c^2$ represents the evolutionary inertia of the dynamical variables. It has dimension of (time)$^2$, i.e. [T]$^2$, in the first dimensional system or [T]$^2$[M]$^{-1}$ in the second dimensional system. $L$ is a quantization parameter, whose dimension is [T] or [T][M]$^{-1}$ in two dimensional systems respectively. The parameter $\beta$ plays the role of $N$-restraining potential whose dimension is zero or [M]$^{-1}$ in two dimensional systems. It is interesting to note that $\frac{L}{c}\beta^{3/2}$ is a dimensionless parameter in both dimensional systems. From Eq (4) we find the parameter is approximately proportional to the relative interval of adjacent levels $\frac{E_{n+1}-E_n}{E_n}$, giving a measure of the quantum peculiarity of the genome. In fact, the parameter takes larger values for angiosperms than other species, making quantum pattern more evident for these plants. For quantum excited states (n>0) of angiosperms that have been studied $\frac{L}{c}\beta^{3/2}$ takes a value between 0.03 and 0.3. However, for quantum singlets (*n*=0) the parameter is lower than above by an order-of-magnitude (Table 3).

**Methods**

By using Hamiltonian **H**, Eq(1), we have succeeded in deriving the discontinuous variation of DNA amount in plant species. The basic equation (2) is the cornerstone of the theoretical analysis that is closely related to the genome evolution.

The genome evolution is governed by a set of constraints fundamental to the complex organization of the system. For eukaryotes the genome interaction between different species is relatively weak and the evolution can be looked as independent genomic events of single species moving on its own trajectory. Consider the problem of eukaryote genome macro-evolution. Suppose the frequency of aptamer (*k*-mer) serves the dynamical variables of the genome (15). Set $\{x_i\}$ ($i=1,...,m=4^k$) are the function of evolutionary time $t$. The genome evolution can be modeled by using different mathematical approaches (23). We model the evolution by a set of differential equations of *k*-mer frequency with respect to *t*. We use the second-order differential equation with respect to *t* instead of first-order equation usually used such as in birth-and-death process modeling. The reason of assuming a second-order equation is: it can easily be transformed to a quantum formation through Feynman's path integral approach (see Supplementary data A, arXiv:1411,2205). Thus, the general form of eukaryote genome evolution is assumed as

$$\frac{d}{dt}(c^2 \frac{dx_i}{dt}) = \frac{\partial V(x_1, x_2,...x_m, t)}{\partial x_i} - f\frac{dx_i}{dt} \qquad (12)$$

where $f > 0$ is a dissipation coefficient representing the effect of fluctuation force. Note that $\sum x_i = N$ (genome length or size). The potential function $V$ (the negative of potential energy) describes the constraints of environment and the interactions among constituents $\{x_i\}$. The parameter $c^2$ is introduced in Eq (12) to represent the evolutionary inertia of the dynamical variables $\{x_i\}$.

To give a unified description of genome evolution both in classical phase and non-classical phase, by the mechanical simulation of Eq (12) we easily obtain its quantum generalization (as $f = 0$)

$$iL\frac{\partial}{\partial t}\psi(\mathbf{x},t) = \mathbf{H}\psi(\mathbf{x},t) \quad (\mathbf{x} = \{x_i\})$$
$$\mathbf{H} = -\frac{L^2}{2c^2}\sum \frac{\partial^2}{\partial x_i^2} - V(\mathbf{x},t) \qquad (13)$$

Here the classical evolutionary trajectory $x_i(t)$ is replaced by the wave function $\psi(\mathbf{x},t)$. $L$ is a quantization constant. The quantum evolutionary equation (13) is the logic generalization of the classical equation (12). Eq (12) has been applied in studying avian genome evolution. (see Supplementary data B, bioRxiv 2015/034710)

Consider the simple case of $k=1$ or $m=4$. Assume the potential

$$V(x_1,...,x_4,t) = D(x_1,...,x_4) + W_{env}(x_1,...,x_4,t) \qquad (14)$$

$D$ means the diversity-promoting potential

$$D(x_1,...,x_4) = N\log_2 N - \sum_i^4 x_i \log_2 x_i, \qquad N = \sum_i^4 x_i \qquad (15)$$

In literature $D$ is called diversity measure which was firstly introduced by Laxton (24). In their studies the geographical distribution of species (the absolute frequencies of the species in different locations) was used as a source of diversity. Recently, the method was developed and applied successfully to various bio-informatics problems, for example, the intron splice site recognition(25), the protein structural classification(26), the nucleosome positioning prediction (27), etc. Now we use it to study evolutionary problem since many examples have shown that the genome always becomes as diverse as possible and expands their own dimensionality continuously in the long term of evolution. Moreover, assume the environmental potential $W_{env}$ in the simple case of $k=1$ is only the function of $N$

$$W_{env} = \alpha N - \beta N^2 \qquad (\beta > 0 \quad N = \sum_i x_i) \qquad (16)$$

Inserting (14), (15), (16) into (13) we obtain

$$iL\frac{\partial}{\partial t}\psi(\mathbf{x},t) = \mathbf{H}\psi(\mathbf{x},t) = \{-\frac{L^2}{2c^2}\sum\frac{\partial^2}{\partial x_i^2} - D(\mathbf{x}) - \alpha N + \beta N^2\}\psi(\mathbf{x},t) \qquad (17)$$

$$D(\mathbf{x}) = N(2-\varepsilon), \quad \varepsilon = 2 + \sum_i \frac{x_i}{N}\log_2\frac{x_i}{N}$$

Set

$$\psi(\mathbf{x},t) = \psi(\mathbf{x})\exp\frac{-iEt}{L}$$

The stationary solution $\psi(\mathbf{x})$ satisfies

$$\{-\frac{L^2}{2c^2}\sum\frac{\partial^2}{\partial x_i^2} - D(\mathbf{x}) - \alpha N + \beta N^2\}\psi(\mathbf{x}) = E\psi(\mathbf{x}) \qquad (18)$$

When $x_i = N/4$ ($\varepsilon = 0$) one has $D(\mathbf{x}) = 2N$. In fact, for many angiosperm plants $\varepsilon$ is a small quantity (for example, $\varepsilon =0.0089$ for Nicotiana) and $D(\mathbf{x}) = 2N$ is a good approximation. In this case, the wave function $\psi$ is a function of variable $N$ only and $\psi(N)$ satisfies Eq (3) exactly. Therefore, we have proved that the basic equation (3), the constraints on genome organization, is essentially the evolutionary equation on the steady states of genome.

**Acknowledgement**

The authors are indebted to Dr Yulai Bao and Ms Aiying Yang for their kind help in literature searching.

**Funding**

This work was supported by National Natural Science Foundation of China [61462068], Inner Mongolia Autonomous Region Natural Science Foundation [2014MS0103] and Award Fund for Special Contributions to Science and Technology of Inner Mongolia Autonomous Region [No 2008 ] .

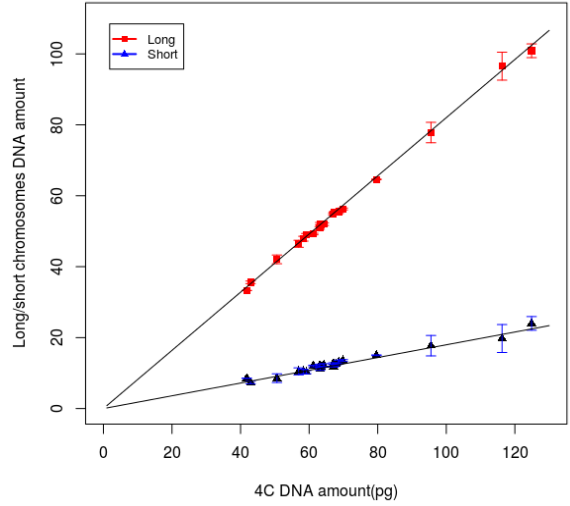

Figure 1　DNA amounts of long and short chromosomes in Aloe

Experimental DNA amounts of long and short chromosomes of 20 Aloe species are plotted by filled square(■) and filled triangle (▲) respectively. Data are taken from (5). The theoretical $N_+$ and $N_-$ versus $N_{tot}$ are calculated from Eq(7) and plotted by the straight lines. The abscissa gives the total 4C DNA amount (pg) and the ordinate is the long or short DNA amount (pg).

Table 1　The nodal distribution of genome size

| Quantum number $n$ | Number of peaks | Equation for determining the positions of peaks | The positions of peaks $\xi_i^{(n)}$ $(i=1,..,n+1)$ |
|---|---|---|---|
| 0 | 1 | $\xi=0$ | 0 |
| 1 | 2 | $\xi^2-1=0$ | 1,-1 |
| 2 | 3 | $2\xi^3-5\xi=0$ | 1.581, 0, -1.581 |
| 3 | 4 | $2\xi^4-9\xi^2+3=0$ | 2.034, 0.602, -0.602, -2.034 |
| 4 | 5 | $4\xi^5-28\xi^3+27\xi=0$ | 2.418, 1.075, 0, -1.075,-2.418 |
| 5 | 6 | $4\xi^6-40\xi^4+75\xi^2-15=0$ | 2.756, 1.475,0.476, -0.476, -1.475, -2.756 |
| 6 | 7 | $8\xi^7-108\xi^5+330\xi^3-195\xi=0$ | 3.063,1.829, 0.882, 0, -0.882, -1.829, -3.063 |
| 7 | 8 | $16\xi^8-280\xi^6+1135\xi^4-1095\xi^2+210=0$ | 3.478, 2.029, 1.017, 0.505, -0.505, -1.017, -2.029, -3.478 |

The peak distribution parameters $\xi$ of genome size are determined by a set of equations listed in third column of the table for ground state $n=0$ and low-lying excited states $n=1$ to 7 respectively. The positions of peaks $\xi_i^{(n)}$ are listed in fourth column in decreasing order for $i=1$ to $n+1$.

Table 2   Nuclear DNA amounts in four genera Clarkia, Nicotiana, Lathyrus and Allium

| Genus | Clarkia | | | | | | | |
|---|---|---|---|---|---|---|---|---|
| $N_i^{(n)}(\exp)$ | 8.87 | 6.13 | 4.51 | 2.42 | | | | |
| $N_i^{(n)}(th1)$ | 8.68 | 6.43 | 4.54 | 2.28 | | | | |
| $N_i^{(n)}(th2)$ | 8.67 | 6.43 | 4.54 | 2.30 | | | | |
| parameter | $N_0$=5.48, $(\sqrt{d})_1$=1.573, $(\sqrt{d})_2$=1.566 | | | | | | | |
| Genus | Nicotiana | | | | | | | |
| $N_i^{(n)}(\exp)$ | 14.57 | 12.09 | 10.68 | 8.20 | 6.22 | 4.67 | 2.86 | |
| $N_i^{(n)}(th1)$ | 14.42 | 12.02 | 10.18 | 8.47 | 6.76 | 4.92 | 2.52 | |
| $N_i^{(n)}(th2)$ | 14.52 | 12.08 | 10.21 | 8.47 | 6.73 | 4.86 | 2.42 | |
| parameter | $N_0$=8.47, $(\sqrt{d})_1$=1.944, $(\sqrt{d})_2$=1.976 | | | | | | | |
| Genus | Lathyrus | | | | | | | |
| $N_i^{(n)}(\exp)$ | 34.20 | 29.21 | 24.53 | 20.51 | 16.88 | 13.80 | 10.76 | 6.86 |
| $N_i^{(n)}(th1)$ | 35.10 | 28.64 | 24.13 | 21.85 | 17.34 | 15.06 | 10.54 | 4.08 |
| $N_i^{(n)}(th2)$ | 34.03 | 28.01 | 23.81 | 21.69 | 17.50 | 15.36 | 11.17 | 5.16 |
| parameter | $N_0$=19.59, $(\sqrt{d})_1$=4.461, $(\sqrt{d})_2$=4.150 | | | | | | | |
| Genus | Allium | | | | | | | |
| $N_i^{(n)}(\exp)$ | 45.90 | 42.72 | — | 34.63 | — | 25.12 | 20.93 | 16.92 |
| $N_i^{(n)}(th1)$ | 41.59 | 37.19 | 34.12 | 32.57 | 29.50 | 27.95 | 24.88 | 20.48 |
| $N_i^{(n)}(th2)$ | 47.21 | 40.63 | 36.04 | 33.72 | 29.13 | 26.81 | 22.22 | 15.64 |
| parameter | $(N_0)_1$=31.04, $(N_0)_2$=31.42, $(\sqrt{d})_1$=3.034, $(\sqrt{d})_2$=4.539 | | | | | | | |

All nuclear DNA amounts listed are given in unit pg. Experimental data $N_i^{(n)}(\exp)$ are group means of 2C DNA taken from (7). Theoretical values $N_i^{(n)}(th1)$ and $N_i^{(n)}(th2)$ are calculated by use of Eq 8 with parameters estimated from Eq 9 and 10 (for $N_i^{(n)}(th1)$) or Eq 9 and 11 (for $N_i^{(n)}(th2)$) respectively. The experimental data were incomplete (namely, for $i$=3, 5) for Allium and the corresponding theoretical calculations $N_i^{(n)}(th2)$ for this genus were obtained by regulating the unknown $N_{i=3}^{(n)}(\exp)$ and $N_{i=5}^{(n)}(\exp)$. The large deviation of $N_i^{(n)}(th1)$ from $N_i^{(n)}(\exp)$ for Allium is due to neglecting the term of $i$=3, 5 in the calculation of square sum of errors.

Table 2a   Standard deviations of different estimates of nuclear DNA amounts
in genera Clarkia, Nicotiana and Lathyrus

| Genus | σ of $N_i^{(n)}(th1)$ | σ of $N_i^{(n)}(th2)$ | σ of equal spacing model |
|---|---|---|---|
| Clarkia | 0.191 | 0.191 | 0.419 |
| Nicotiana | 0.342 | 0.335 | 0.415 |
| Lathyrus | 1.258 | 1.076 | 1.420 |

The standard deviations σ of different estimates of nuclear DNA amount are compared for three genera which have complete experimental data.  The equal spacing model is a purely empirical one.  Define empirical spacing parameter

$$\xi_{emp}^{(n)} = \frac{N_1^{(n)} - N_{n+1}^{(n)}}{n}$$

and predicted $N_i^{(n)}$ (denoted as $N_i^{(n)}(emp)$)

$$N_i^{(n)}(emp) = \{N_1^{(n)}, N_1^{(n)} - \xi_{emp}^{(n)}, N_1^{(n)} - 2\xi_{emp}^{(n)}, ....., N_1^{(n)} - n\xi_{emp}^{(n)} = N_{n+1}^{(n)}\}$$

Table 3  Nuclear quantum state and DNA amounts of species in subgenus Eurosa of Roses

| Section | $n$ | Quantum State $S_i^{(n)}$ | 2C DNA amount (pg ± s.d.) | $N_0$ | $\sqrt{d}$ | $\dfrac{L}{c}\beta^{3/2}$ |
|---|---|---|---|---|---|---|
| Banksianae | 0 | $S_1^{(0)}$ | 1.04±0.08 | 1.04 | 0.11 | 0.0084 |
| Bracteatae | 0 | $S_1^{(0)}$ | 1.15±0.08 | 1.15 | 0.11 | 0.0068 |
| Caninae | 0 | $S_1^{(0)}$ | 2.91±0.08 | 2.91 | 0.11 | 0.0031 |
| Carolinae | 1 | $S_2^{(1)}$ | 0.95±0.08 | 1.45 | 0.50 | 0.084 |
|  |  | $S_1^{(1)}$ | 1.95±0.08 |  |  |  |
| Cinamomeae | 2 | $S_3^{(2)}$ | 0.95±0.08 | 1.95 | 0.64 | 0.076 |
|  |  | $S_2^{(2)}$ | 1.93±0.08 |  |  |  |
|  |  | $S_1^{(2)}$ | 2.96±0.09 |  |  |  |
| Gallicanae | 0 | $S_1^{(0)}$ | 2.20±0.08 | 2.20 | 0.11 | 0.0019 |
| Indicae | 0 | $S_1^{(0)}$ | 1.16±0.08 | 1.16 | 0.11 | 0.0067 |
| Laevigatae | 0 | $S_1^{(0)}$ | 1.14±0.08 | 1.14 | 0.11 | 0.0069 |
| Pimpinellifoliae | 1 | $S_2^{(1)}$ | 0.78±0.08 | 1.34 | 0.56 | 0.124 |
|  |  | $S_1^{(1)}$ | 1.90±0.08 |  |  |  |
| Synstylae | 0 | $S_1^{(0)}$ | 1.13±0.08 | 1.13 | 0.11 | 0.0071 |

The nuclear quantum state of each section in the subgenus Eurosa of Roses is defined and denoted as $S_i^{(n)}$ in the table. The 2C DNA amount is also listed which was taken from the data given in ref 8 (8). Seven sections fall into the genome of ground state denoted by $n=0$ with one value of 2C DNA amount. Two sections fall into the genome of first excited state denoted by $n=1$ with two discrete values of 2C DNA amount. One section falls into the genome of second excited state denoted by $n=2$ with three discrete values of 2C DNA amount. $N_0$ and $\sqrt{d}$ are calculated from Eq (9) and (10) and $\sqrt{d}$ for ground states are estimated from the standard deviation of 2C DNA amount.